\documentclass{aa}
\def\simlt{\lower.5ex\hbox{$\; \buildrel < \over \sim \;$}}
\def\simgt{\lower.5ex\hbox{$\; \buildrel > \over \sim \;$}}
\begin{document}

   \thesaurus{01     
              (02.01.2;  
	       02.02.1;  
	       08.02.3;
               08.05.1;  
	       08.23.3;  
               13.25.3)}  
   \title{XMM-Newton Optical Monitor observations of LMC X-3}


   \author{R. Soria\inst{1}, K. Wu
          \inst{1}\fnmsep\inst{2}, M. J. Page\inst{1}
          \and
          I. Sakelliou\inst{1}
          }

   \offprints{R. Soria}

   \institute{Mullard Space Science Laboratory, University College London,
              Holmbury St Mary, Dorking, RH5 6NT, UK\\
              email: rs1@mssl.ucl.ac.uk
         \and
             School of Physics, University of Sydney, Sydney 2006, NSW, 
             Australia\\
             }

   \date{Received XXXX; accepted XXXX}

   \authorrunning{R. Soria, K. Wu, M. Page \& I. Sakelliou}

   \maketitle

   \begin{abstract}

   We study the optical counterpart of the black-hole 
X-ray binary LMC X$-$3, by using 
{\it XMM-Newton/OM} observations carried 
out during a low-hard X-ray state. 
We derive a better constraint for the temperature, mass 
and radius of the companion star, and we show that the   
star is likely to be a $\sim$ B5 subgiant filling 
its Roche lobe. Taking into account the effect of X-ray irradiation, 
we suggest a value $f_{\rm M} \approx 1.5 M_{\odot}$ for the 
mass function in this system, lower than previously thought; we 
provide a more accurate lower limit to the mass of the compact object.

   \end{abstract}

%

\section{Introduction}

An important unsolved problem for the black-hole candidate (BHC) LMC X$-$3 
is the process of mass transfer. With an estimated
mass of the companion star $4 M_{\odot} \simlt M_2 \simlt 8 M_{\odot}$ 
(Cowley et al.\ 1983), 
the system appears to be intermediate between high-mass black-hole binaries 
like Cyg X$-$1 (mass of the companion star $\approx 33 M_{\odot}$, 
see Giles \& Bolton 1986), 
and low-mass black-hole binaries like A0620$-$00 
(mass of the companion star $\approx 0.7 M_{\odot}$). 
In the former class of systems, 
mass transfer occurs mainly via a stellar wind, and the donor star 
is more massive than the primary; in the latter, the donor star 
is usually a late-type star filling its Roche lobe.

The optical counterpart (Warren and Penfold 1975) of LMC X$-$3 
shows ellipsoidal modulations with a total amplitude 
of $\simlt 0.2$ mag and a period corresponding 
to the 1.705d binary period (van der Klis, Tjemkes and van Paradijs 1983; 
van der Klis et al 1985). It also shows long-term 
brightness variations in the range $16.7 \simlt V \simlt 17.5$
(e.g., van Paradijs et al.\ 1987). Its optical spectrum 
was found to be consistent with that of a B3 main sequence star (Cowley 
et al.\ 1983). From the radial velocity shifts of the optical 
absorption lines, Cowley et al.\ (1983) derived a mass function  
$f_{\rm M} = 2.3 M_{\odot}$, thus establishing this system as a strong BHC. 
However, the spectral identification of the companion star 
remains uncertain (Mazeh et al.\ 1986) because of the 
effect of X-ray irradiation on the secondary, and the possible
contribution to the optical flux by an accretion disk.
In fact, the long-term variations in 
both the optical colours and the $V$ brightness 
are found to be 
associated with changes in the soft X-ray flux (Cowley et al.\ 1991).

Moreover,
Nowak et al.\ (2000) and Wilms et al.\ (2000) suggest that 
spectral transitions 
between a high-soft and a low-hard state could be associated with 
changes in the accretion rate.  A more correct identification of the 
companion star can provide a better understanding of the physics of 
mass transfer and state transitions.

\section{{\it{XMM-Newton}} Optical Monitor observations}

LMC X$-$3 was observed with the Optical Monitor 30in telescope 
(Mason et al.\ 1996) 
on board {\it{XMM-Newton}} on 2000 April 19 
(MJD 51653; {\it{XMM-Newton}} revolution 66). 
A log of the observations is reported in Table 1.
The system was in a low-hard state during our optical observation, 
with a pure power-law X-ray spectrum with photon index $\approx 1.6$--$1.8$ 
(Boyd and Smale 2000; Wu et al.\ 2000). 
We estimate from simultaneous {\it{XMM-Newton/PN}} observations 
(Wu et al.\ 2000) that the 2--10 keV flux was $\simlt 10^{36}$ erg s$^{-1}$, 
at least two orders of magnitude lower than in the high-soft state 
(Nowak et al.\ 2000; Wilms et al.\ 2000). 
This is in agreement with the RXTE/ASM 
2--10 keV lightcurve. Therefore, optical photometric observations 
of the companion star at this epoch provide the best chance 
to determine its intrinsic colours and spectral type.

\section{Optical colours and temperature}

The average brightness of LMC X$-$3 measured 
in the three {\it{XMM-Newton/OM}} optical bands on 2000 April 19 was 
$v = 17.48 \pm 0.02$, $b = 17.39 \pm 0.02$, $u = 16.56 \pm 0.02$. 
(For a detailed comparison between the {\it{XMM-Newton/OM}} 
photometric system and Johnson's standard $UBV$ system, see
Royer et al.\ (2000)). Using the latest available matrix of colour 
transformation coefficients (Version 4, 2000 September), 
we find that this corresponds to 
$V = 17.47 \pm 0.03$, $B = 17.35 \pm 0.03$, $U = 16.77 \pm 0.03$.

Our observations covered a fraction 
of the binary phase; therefore they may not yield the true 
phase-averaged brightness of the optical counterpart.
If we adopt the ephemeris of 
van der Klis (1985), the $V$ images were centred at phase 
$\phi = 0.86 \pm 0.10$, the $B$ images at $\phi = 0.95 \pm 0.10$ 
and the $U$ images at $\phi = 0.90 \pm 0.10$.
However, those ephemeris may have accumulated too large 
an error to be still reliable. From the amplitude of the ellipsoidal 
variations, we estimate that there is
an additional error of 0.08 mag in the $V$ brightness,
of 0.03 mag in the $U-B$ colour, and of 0.04 mag in the $B-V$ colour.

\begin{figure}
\begin{center}
\leavevmode
\setlength{\unitlength}{1cm}
\begin{picture}(8.8,6)
\put(0,0){\includegraphics{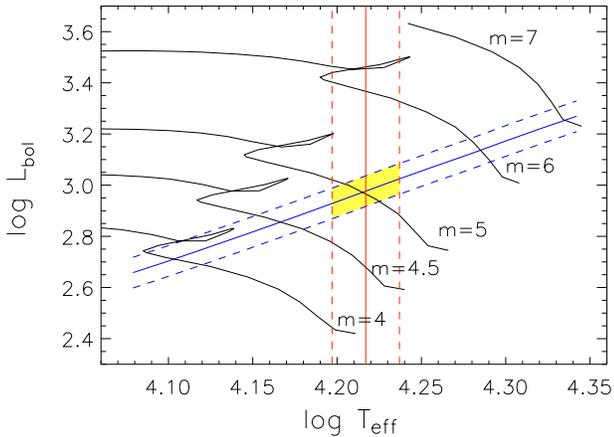}}
\end{picture}
\caption{Evolutionary tracks for stars of various masses, 
at Z = 0.008, in the 
($\log T_{\rm eff}$,\,$\log L_{\rm bol}$) plane, with the acceptable 
range of temperatures and luminosities derived from our {\it{XMM-Newton/OM}} 
observations. Masses are in units of solar mass,
$M_{\odot} = 1.99 \times 10^{33}$ g; 
temperature is in K; luminosity is in units of solar 
bolometric luminosity, $L_{\rm bol, \odot} = 3.9 \times 10^{33}$ 
erg s$^{-1}$. 
}
\label{fig:epicspec}
\end{center}
\end{figure}

If we assume a reddening $E(B-V) = 0.06 \pm 0.01$ (Schlegel, 
Finkbeiner \& Davis 1998; 
Crampton 1979)
and a distance modulus $d = 18.5 \pm 0.1$, 
we obtain intrinsic colours $(B-V)_0 = -0.17 \pm 0.05$, 
$(U-B)_0 = -0.64 \pm 0.04$, and an absolute magnitude 
$M_V = -1.22 \pm 0.16$ for the optical counterpart of LMC X$-$3.

Both optical colour indices can be used as indicators of the effective 
temperature $T_{\rm eff}$. From Buser and Kurucz (1978), and 
Cramer (1984), we find that these intrinsic colours 
correspond to stars with $15500 \simlt T_{\rm eff} 
\simlt 17000$ K. The colour--temperature relation is almost independent 
of surface gravity, so that it can be applied to both 
main sequence and giant stars. 
We also checked the effect of different metal abundances by using 
the semi-empirical colour-temperature calibration of 
Lejeune, Cuisinier and Buser (1998). For [Fe/H]$ = -0.3$, our observed 
colour indices imply $15600 \simlt T_{\rm eff} \simlt 17200$, 
while for [Fe/H]$ = -1.5$ they imply $16300 \simlt T_{\rm eff} \simlt 18000$. 
If we adopt the slightly bluer colours for the companion star 
measured by van der Klis et al.\ (1983), we obtain temperatures 
$T_{\rm eff} \approx 18000$--$19000$.

From the catalogue of Lejeune et al.\ (1998) we derive the 
bolometric corrections, which depend on the temperature and (weakly) 
on the metal abundance, but are almost independent of surface gravity.
From the inferred range of bolometric luminosities and temperatures 
we obtain a reliable constraint to the mass 
of the secondary star. 

Figure 1 shows the single-star evolutionary tracks 
from Girardi et al.\ (2000) in the ($\log T_{\rm eff}$,\,$\log L_{\rm bol}$) 
plane, for a metal abundance $Z = 0.008$ (a typical value for the 
LMC, see for example Caputo, Marconi and Ripepi 1999). 
Assuming that evolutionary tracks for stars in binary systems 
can be approximated by single-star tracks, we infer that 
the constraint is satisfied 
by evolved stars with $4.7 M_{\odot} \simlt M_2 \simlt 5.3 M_{\odot}$ 
corresponding to subgiants of spectral type $\sim$ B5. 
A B3 main sequence companion ($M_2 \approx 6$--7$M_{\odot}$) has
the correct luminosity but much higher temperatures.
This result is not very sensitive to the adopted value of metal abundance.

\begin{table}
\caption{Log of our {\it{XMM-Newton/OM}} observations}
\begin{tabular}{ccc}
\hline
OM filter & Mid-exposure time (MJD) & Exposure time (s)\\
\hline
 & &\\
v & 51653.148 & 1000\\
v & 51653.178 & 1000\\
v & 51653.193 & 1000\\ 
v & 51653.208 & 1000\\
u & 51653.123 & 1000\\
u & 51653.238 & 1000\\
u & 51653.253 & 1000\\
u & 51653.268 & 1000\\
u & 51653.283 & 1000\\
b & 51653.298 & 1000\\
b & 51653.313 & 1000\\
b & 51653.328 & 1000\\
b & 51653.343 & 1000\\
b & 51653.358 & 1000\\
\end{tabular}
\end{table}

\section{Radius of the companion star}

The mean mass density in the Roche lobe of the companion star 
is uniquely determined by the orbital period $P$
(Frank, King and Raine 1992):
\begin{equation}
\rho \equiv \frac{3M_2}{4 \pi R_{\rm L}^3} \approx 115 P^{-2}_{\rm hr} 
\approx 0.069 \ \rm{g\  cm}^{-3},
\end{equation}
where $R_{\rm L}$ is the mean radius of the Roche lobe.  
We plot in Figure 2 the evolutionary tracks (Girardi et al.\ 2000)
in the ($M_V$,\,$\rho$) plane, together with the mean density 
in the Roche lobe derived in Equation (1), for Z = 0.008. The dashed line 
corresponds to a radius of $0.95 R_{\rm L}$.
Stars with $4 M_{\odot} \simlt M_2 \simlt 4.7 M_{\odot}$ 
would be very close to filling their Roche lobe at the observed 
brightness $M_V$.
If the companion star has a mass $M_2 = 4.7 M_{\odot}$, 
its radius would be 
$R_2 = 4.4 R_{\odot} \simeq R_{\rm L}$. Mass transfer 
would mainly occur via Roche-lobe overflow. 
Stars of lower mass would also fill their Roche lobe but are ruled out 
by the observed range of colours (Figure 1). On the other hand, 
if $M_2 > 4.7 M_{\odot}$, the companion star would not fill its Roche 
lobe. In particular, a B3V companion ($R_2 \approx 4 R_{\odot}$) 
would only fill less than half of the volume of its 
Roche lobe. The main mechanism of mass transfer would have to be 
a stellar wind. If the wind were to account 
for X-ray luminosities up to $\approx 4 \times 10^{38}$ erg s$^{-1}$ 
in the high-soft 
state (Johnston, Bradt and Doxsey 1979), the column density
would be much higher than 10$^{21}$ cm$^{-2}$, difficult to reconcile 
with the result obtained from 
the {\it{XMM-Newton}} X-ray observations (Wu et al.\ 2000).

\begin{figure}
\begin{center}
\leavevmode
\setlength{\unitlength}{1cm}
\begin{picture}(8.8,6)
\put(0,0){\includegraphics{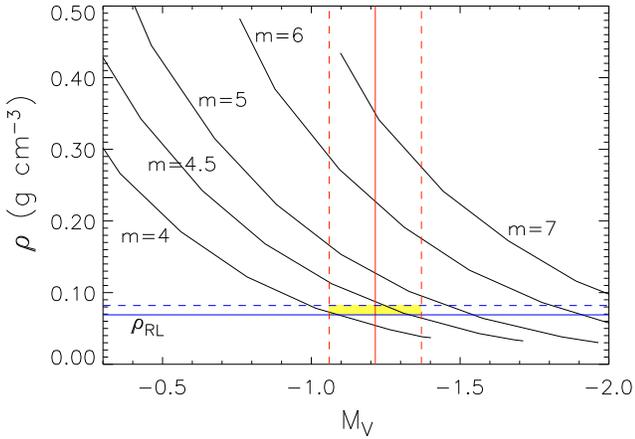}}
\end{picture}
\caption{Evolutionary tracks for stars of various masses, 
at Z = 0.008, in the ($M_V$,\,$\rho$)
plane, compared with the 
absolute $V$ brightness derived from our {\it{XMM-Newton/OM}} 
observations, and the mean density inside the Roche lobe 
of the secondary star. The dashed line corresponds to a radius 
$R = 0.95 R_{\rm L}$ for which Roche-lobe overflow becomes significant. 
}
\label{fig:epicspec}
\end{center}
\end{figure}

The evolutionary time-scale for a subgiant star of mass 
$\approx 4.7 M_{\odot}$ in the observed range of temperature 
is $\approx 2 \times 10^7$ years 
(Girardi et al.\ 2000). If the mass transfer is driven by the nuclear 
evolution of the secondary, this would imply a mass-transfer rate 
$\approx 2 \times 10^{-7} M_{\odot}$ per year, sufficient to account for 
luminosities $\simgt 10^{39}$ erg s$^{-1}$.

\section{A lower limit on the mass of the primary}

The masses of the two components of a binary system 
are given by
\begin{equation}
M_{\rm X} = \frac{(1+q)^2}{\sin^3 i}f_{\rm M},
\end{equation}
\begin{equation}
M_{\rm 2} = \frac{q(1+q)^2}{\sin^3 i}f_{\rm M},
\end{equation}
where $q \equiv M_2/M_{\rm X}$ and $f_{\rm M}$ is the mass function 
of the primary, which can be obtained from spectroscopic observations.
The X-rays are not eclipsed (Cowley et al.\ 1983); 
if the secondary star is approximated by a sphere, this implies 
(Paczy\'{n}ski 1983):
\begin{equation}
R_2 < K_2 \frac{P}{2\pi}(1+q) \cot i,
\end{equation}
where $K_2$ is the projected radial velocity semi-amplitude of the 
secondary star.
From equations (3) and (4), an upper limit for $q$ is obtained:
\begin{eqnarray}
R_2 < & K_2 P (2\pi)^{-1} (1+q)^{1/3} q^{-1/3} f_{\rm M}^{-1/3}M_2^{1/3} \times 
\nonumber \\ 
 & \left[1-q^{2/3}(1+q)^{4/3}
f_{\rm M}^{2/3}M_2^{-2/3}\right]^{1/2},
\end{eqnarray}
where the right-hand side is a monotonically decreasing function of $q$ 
where defined.

From the radial velocity shifts of the stellar absorption lines, 
Cowley et al.\ (1983) determined $K_2 = 235 \pm 11$ km s$^{-1}$ and 
$f_{\rm M} = 2.3 \pm 0.3 M_{\odot}$. Using these values together with 
the values of $M_2$ and $R_2$ determined in Section 4, we obtain 
from equation (5):
\begin{equation}
M_{\rm X} > 7.3 \pm 0.6 M_{\odot}.
\end{equation}
As an aside, we also note that if we assume a lower limit $i > 50$ 
for the inclination angle (van der Klis et al.\ 1985), equation (2) 
gives an upper limit $M_{\rm X} < 13.8 \pm 1.0 M_{\odot}$.


If the secondary star is really a B5 subgiant, 
its intrinsic luminosity 
is $\simlt 3.5 \times 10^{36}$ erg s$^{-1}$ (Figure 1), corresponding 
to an intrinsic flux $F_2 \simlt 3 \times 10^{12}$ erg cm$^{-2}$ s$^{-1}$. 
The radial velocity curve of the stellar 
absorption lines was obtained by Cowley et al.\ (1983) when the source 
was in a high state, implying an X-ray luminosity $L_{\rm X} \simgt 10^{38}$ 
erg s$^{-1}$. The orbital separation in a binary system is 
$a = 2.9 \times 10^{11} M_{1}^{1/3}(1+q)^{1/3}P_{\rm d}^{2/3} 
\approx 1.0 \times 10^{12}$ cm for LMC X$-$3.
Hence, the energy intercepted by the secondary star per unit time 
is $(1/4)L_{\rm X}(R_2/a^2)^2 \simgt 3 \times 10^{36}$ erg s$^{-1}$, 
comparable with the intrinsic luminosity of the star.

In a spherical approximations, the intercepted 
flux is larger than the intrinsic flux on $\approx 1/4$ of the 
surface of the star, for a soft X-ray luminosity $\approx 2 \times 10^{38}$ 
erg s$^{-1}$, and for a disk with $H/R \simlt 0.2$.
Strong external irradiation with soft X-rays 
tends to cause photo-ionisation at the surface of the secondary 
star. As a consequence, absorption lines from the irradiated 
face should be weakened or suppressed, and the observed radial velocity shifts 
may not reflect the true orbital motion of the centre of mass 
of the secondary (Wade and Horne 1988; Phillips, Shahbaz 
and Podsiadlowski 1999). 

Using the model previously applied to the BHC GRO J1655$-$40 
(Phillips, Shahbaz and Podsiadlowski 1999), and 
assuming for simplicity 
that no absorption lines are 
produced in the region where the external flux is larger than the 
internal flux, we can estimate the effect of irradiation 
in the case of LMC X$-$3.
The observed amplitude of the radial velocity variations
of the absorption lines ($K_2$) should appear larger than the true radial 
velocity amplitude of the centre of mass of the secondary 
($K_{2}' \equiv K_2 - \Delta K_2$). 
For the parameters of LMC X$-$3, we estimate that the correction is
$\Delta K_2 = 30 \pm 5$ km s$^{-1}$. Therefore, 
the true $K$-velocity of the secondary star is only 
\begin{equation}
K_{2}' = 205 \pm 12 {\rm \ km\ s}^{-1}.
\end{equation}

Hence, after the correction, the mass function ($\propto K_2^3$) 
for the compact object in LMC X$-$3 is 
\begin{equation}
f_{\rm M} = 1.5 \pm 0.3 M_{\odot}.
\end{equation}
By using these revised $K$-velocity and mass function in 
equation (5), we obtain a lower limit to the mass of the 
compact object: 
\begin{equation}
M_{\rm X} > 5.8 \pm 0.6 M_{\odot}.
\end{equation} 

It would be important to test this prediction by determining 
the radial velocity curve of the stellar lines during a low X-ray state.
We also note that if the irradiated side of the 
secondary star contributes only to the optical continuum flux 
but not to the absorption lines, their relative strength would 
appear reduced in comparison with a non-irradiated star of the same 
spectral type. This is a possible explanation of what 
was observed by Cowley et al.\ (1983).

\section{Conclusions}

We have observed the BHC LMC X$-$3 with the Optical Monitor on board 
{\it{XMM-Newton}} during an X-ray low-hard state. The 
$2$--$10$ keV flux at that epoch 
was the lowest ever observed by RXTE/ASM, consistent with zero.
The brightness and colours of the optical counterpart inferred 
from our observations are likely to be a good approximation 
of the intrinsic values of the companion star. This allows us 
to constrain the mass and radius of the secondary. We have found that 
an evolved subgiant of mass $M_2 \approx 4.7 M_{\odot}$ and 
temperature $T_{\rm eff} \approx 16500$ K (spectral type $\sim$ B5\,IV)
is consistent with the observed colours and luminosity. 
We have also shown that such a star would be filling its Roche lobe, 
thus explaining why the source is mostly observed in a high-soft state,
dominated by an accretion disk. No significant wind is expected from such 
star, in agreement with the low column density inferred 
from the {\it XMM-Newton/EPIC} and {\it RGS} X-ray data. 
The evolutionary time-scale of a B5\,IV star is also consistent with 
the observed X-ray luminosity powered by nuclear-evolution driven 
mass transfer.

The companion star was previously thought to be a main-sequence B3 star. 
Although it would have the same brightness, a B3\,V star would fill 
only about one half of its Roche lobe, ruling out mass transfer 
via Roche-lobe overflow.

A B5\,IV companion implies a lower 
limit to the mass of the primary $M_{\rm X} > 7.3 \pm 0.6 M_{\odot}$, 
if we assume the $K$-velocity and mass function 
determined by Cowley et al.\ (1983).
However, the spectroscopic observations 
of Cowley et al.\ (1983) were carried out during an X-ray active 
state, when the effect of soft X-ray irradiation 
on the surface of the secondary star is significant. 
The true radial velocity amplitude of the 
centre of mass of the secondary is probably 
$\approx 12$--$15$ per cent less than 
the value observed by Cowley et al.\ (1983). This would reduce 
the mass function to $f_{\rm M} = 1.5 \pm 0.3 M_{\sun}$, and set a lower 
limit $M_{\rm X} > 5.8 \pm 0.6 M_{\odot}$ for the mass of the 
compact object.

\begin{acknowledgements}
 
  KW acknowledges the support from the ARC Australian Research Fellowship 
     and a PPARC visiting fellowship. We thank Keith Mason for his 
     comments.

\end{acknowledgements}


\begin{thebibliography}{}
 

   \bibitem[]{}Boyd, P. T., Smale, A. P., 2000, IAUC 7424

   \bibitem[]{}Buser, R., Kurucz, R. L., 1978, A\&A 70, 555

   \bibitem[]{}Caputo, M., Marconi, G.,  Ripepi, V., 1999, ApJ 525, 784

   \bibitem[]{}Cowley, A. P., Crampton, D., Hutchings, J. B., 
	Remillard, R., Penfold, J. E. 1983, ApJ 272, 118

   \bibitem[]{}Cowley, A. P., Schmidtke, P. C., Ebisawa, K., Makino, F.,
 	Remillard, R. A., Crampton, D., Hutchings, J. B., Kitamoto, S.,
 	Treves, A., 1991, ApJ 381, 526

   \bibitem[]{}Cramer, N., 1984, A\&A 132, 283

   \bibitem[]{}Crampton, D., 1979, ApJ 230, 717

   \bibitem[]{}Frank, J., King, A., Raine, D. 1992, Accretion Power 
	in Astrophysics (Cambridge: University Press)

   \bibitem[]{}Giles, D. R., Bolton, C. T., 1986, ApJ 304, 37

   \bibitem[]{}Girardi, L., Bressan, A., Bertelli, G., Chiosi, C., 2000, 
	A\&AS 141, 371

   \bibitem[]{}Johnston, M. D., Bradt, H. V., Doxsey, R. E., 1979, 
	ApJ 233, 514

   \bibitem[]{}Lejeune, T., Cuisinier, F., Buser, R., 1998, A\&AS 130, 65

   \bibitem[]{}Mason, K. O., Cropper, M. S., Hunt, R., Horner, S. D., 
	Priedhorsky, W. C., Ho, C., Cordova, F. A., Jamar, C. A., 
	Antonello, E., 1986, Proc. SPIE Vol.\ 2808 (eds: 
	O. H. Siegmund, M. A. Gummin), 438

   \bibitem[]{}Mazeh, T., van Paradijs, J., van den Heuvel, E. P. J., 
	Savonije, G. J., 1986, A\&A 157, 113

   \bibitem[]{}Nowak, M. A., Wilms, J., Heindl, W. A., Pottschmidt, K., 
	Dove, J. B., Begelman, M. C., 2000, MNRAS, in press 
	(astro-ph/0005487)

   \bibitem[]{}Paczy\'{n}ski, B. 1983, ApJ 273, L81

   \bibitem[]{}Phillips, S. N., Shahbaz, T., Podsiadlowski, Ph., 1999,
	MNRAS 304, 839

   \bibitem[]{}Royer, P., Manfroid, J., Gosset, E., Vreux, J.-M., 2000, 
	A\&AS 145, 351

   \bibitem[]{}Schlegel, D. J., Finkbeiner, D. P., Davis, M., ApJ 500, 525

   \bibitem[]{}van der Klis, M., Tjemkes, S., 
	van Paradijs, J., 1983, A\&A 126, 265 
	(see also {\it Erratum} 132, 240 (1984))

   \bibitem[]{}van der Klis, M., Clausen, J. V., Jensen, K., Tjemkes, S., 
	van Paradijs, J., 1985, A\&A 151, 322

   \bibitem[]{}van Paradijs, J., van der Klis, M.,
 	Augusteijn, T., Charles, P., Corbet, R. H. D., 
	Ilovaisky, S., Maraschi, L., Motch, C., Pakull, M.,
 	Smale, A. P., Treves, A., van Amerongen, S., 1987, A\&A 184, 201

   \bibitem[]{}Wade, R. A., Horne, K., 1988, ApJ 324, 411

   \bibitem[]{}Warren, P. R., Penfold, J. E., 1975, MNRAS 172, 41P

   \bibitem[]{}Wilms, J., Nowak, M. A., Pottschmidt, K., Heindl, W. A., 
	Dove, J. B., Begelman, M. C., 2000, MNRAS, in press
	(astro-ph/0005489)

   \bibitem[]{}Wu, K., Soria, R., Page, M. J., Sakelliou, I., 2000, 
	submitted to A\&A


\end{thebibliography}
\end{document}